# Imaging anisotropic waveguide exciton polaritons in tin sulfide

Yilong Luan[1,2], Hamidreza Zobeiri[3], Xinwei Wang[3], Eli Sutter[4,5], Peter Sutter[6]*, Zhe Fei[1,2]*

[1]Department of Physics and Astronomy, Iowa State University, Ames, Iowa 50011, USA
[2]Ames Laboratory, U. S. Department of Energy, Iowa State University, Ames, Iowa 50011, USA
[3]Department of Mechanical Engineering, Iowa State University, Ames, IA 50011, USA
[4]Department of Mechanical and Materials Engineering, University of Nebraska-Lincoln, Lincoln, NE 68588, USA
[5]Nebraska Center for Materials and Nanoscience, University of Nebraska-Lincoln, Lincoln, NE 68588, USA.
[6]Department of Electrical and Computer Engineering, University of Nebraska-Lincoln, Lincoln, NE 68588, USA

*Corresponding to: (P.S.) psutter@unl.edu, (Z.F.) zfei@iastate.edu.

**Abstract**

In recent years, novel materials supporting in-plane anisotropic polaritons have attracted a lot of research interest due to their capability of shaping nanoscale field distributions and controlling nanophotonic energy flows. Here we report a nano-optical imaging study of waveguide exciton polaritons (EPs) in tin sulfide (SnS) in the near-infrared (IR) region using the scattering-type scanning near-field optical microscopy (s-SNOM). With s-SNOM, we mapped in real space the propagative EPs in SnS, which show sensitive dependence on the excitation energy and sample thickness. Moreover, we found that both the polariton wavelength and propagation length are anisotropic in the sample plane. In particular, in a narrow spectral range from 1.32 to 1.44 eV, the EPs demonstrate quasi-one-dimensional propagation, which is rarely seen in natural polaritonic materials. Further analysis indicates that the observed polariton anisotropy is originated from the different optical bandgaps and exciton binding energies along the two principal crystal axes of SnS.

**Main text**

In-plane anisotropic polaritons[1,2] were first studied in metasurfaces[3-5] where nanostructuring of the polaritonic media or substrates breaks the symmetry, thus enabling polaritonic anisotropy. Later, several natural materials were predicted and/or experimentally confirmed to support in-plane anisotropic polaritons.[6-10] For example, anisotropic plasmon polaritons and hybrid plasmon-phonon polaritons were observed in black phosphorus carbides with far-field infrared (IR) spectroscopy.[8] Anisotropic phonon polaritons with hyperbolic wavefronts were imaged in $MoO_3$,[9-11] which can be conveniently tailored by controlling the sample thickness and by stack- and twist-engineering.[12-18] Compared to nano-engineered anisotropic metasurfaces, natural materials with intrinsic anisotropic polaritons are generally more convenient for applications and can avoid potential material quality degradation due to complex nano-fabrications. Despite these advantages, natural materials supporting in-plane anisotropic polaritons are rare and are so far mainly studied in the mid-IR range. New materials enabling anisotropic polaritons in other technologically important spectral regions (e.g., near-IR and visible) are desired.

In this Letter, we report the experimental discovery of strongly-anisotropic exciton polaritons (EPs) in tin sulfide (SnS) in the technologically-important near-IR region. SnS is a post-transition-metal monochalcogenide and a van der Waals (vdW) layered semiconductor with an orthorhombic structure, analogous to that of black phosphorous.[6-7] As sketched in Figure 1b, the two in-plane axes of SnS, namely the *a* and *b* axes, are along the zigzag and armchair directions, respectively. SnS has been widely studied due to its unique anisotropic optoelectronic properties[19-23] and potential applications related to photodetection and solar energy harvesting.[24-27] In particular, the energies of excitons or optical bandgaps along the *a* and *b* axes of SnS are about $E_a \approx 1.39$ eV and $E_b \approx 1.66$ eV respectively,[22,23] which directly impact the polaritonic responses. Note that EPs have previously been studied in other vdW semiconductors (e.g. WSe$_2$, MoSe$_2$, etc.) with imaging[28-32] and spectroscopic methods,[33-37] where the EPs are isotropic in the sample plane. The samples studied here are SnS microcrystals supported on mica wafers (Figure S1a). As introduced in detail in the earlier work,[19] these microcrystals have a wrap-around layered core-shell structure: the thick SnS core is coated with a thin crystalline shell (thickness $\approx 3$ nm) of layered tin disulfide (SnS$_2$). A detailed characterization of the wrap-around core-shell structures and their synthesis process were reported in the earlier work.[19] Note that the thin SnS shell is isotropic in the sample plane[38,39], so the observed anisotropic properties of EPs are solely due to the SnS core. The SnS$_2$ shell mainly serves as a protection layer of the SnS core and the waveguide EPs. Detailed discussions about the effect of the SnS$_2$ shell are given in the Supporting Information.

To excite and probe EPs in SnS, we employed a scattering-type scanning near-field optical microscope (s-SNOM) that was built based on an atomic force microscope (AFM). As illustrated in Figure 1a, the sharp metalized tip in s-SNOM excited by a *p*-polarized laser beam generates strong evanescent fields underneath the tip. These evanescent fields with a wide range of wavevectors[40] can effectively excite transverse-magnetic (TM) polaritons inside the sample.[29] The excitation source used in the study is a broadband (1.24-1.77 eV) Ti:sapphire laser that covers the bandgap and exciton energies of SnS (see discussions below). We used a parabolic mirror to focus the laser beam at the tip apex, and the scattered photons off the tip/sample system are collected by the same parabolic mirror and then counted by a photodetector. More introductions about the nano-optical setup are in Supporting Information.

In Figure 1c, we plot the AFM topography image of a typical SnS microcrystal coated with a thin SnS$_2$ shell.[19] The lateral sizes of the crystal are approximately 6-8 μm, and the thickness is about 100 nm including the SnS$_2$ shell. Here the crystal has a total of eight edges, among which the four short edges are along the *a* or *b* axes.[19] We were able to determine the crystal axes of the sample by examining the shape of the crystal and by Raman spectroscopy (see Supporting Information). Figure 1d plots the near-field amplitude (*s*) images taken simultaneously with the topography image (Figure 1c) at the excitation energy of $E = 1.38$ eV. Here, the in-plane wavevector of the laser ($k_{in}$) is along the *b* axis of SnS. From Figure 1d, one can see many interference fringes and oscillations inside the sample. We focus on a string of one-dimensional (1D) oscillations extending from the left edge to the crystal center along the *b* axis (marked with a white arrow). According to previous studies,[29,30] these oscillations are generated due to the interference between two major beam paths as sketched in Figure 1a. In the first path (P1), the excitation photons are scattered back directly by the tip apex. In the second path (P2), the excitation photons are first transferred into waveguide EPs by the s-SNOM tip. These EPs then propagate toward the sample edge and get scattered into photons. Photons collected through the two beam paths are coherent with each other, so they can generate interference. When scanning the tip perpendicular to the sample edge, the distance between the tip and the sample edge varies, so a string of bright and dark spots forms due to constructive and destructive interferences, respectively. As sketched in Figure 1a, the left short edge of the crystal is responsible for the generation of the 1D interference oscillations along the direction of the white arrow in Figure 1d. Other edges can also scatter EPs into photons and generate interference patterns. For example, the four long edges that are about 43° relative to the *b* axis are responsible for the bight fringes parallel to these edges (see Figure S2b). There are other possible interference mechanisms (e.g., edge excitation of polaritons), but they are not responsible for the fringes/oscillations observed in our samples. Detailed discussions of different interference mechanisms are given in Section 3 of the Supporting Information. From Fig. 1d,f, we also seen fringes on the substrate side, which are generated due to the excitation and scattering of photons at the air/mica interface as confirmed by dispersion analysis (see Figure S10).

Figure 1e,f plot the AFM and corresponding s-SNOM imaging data of the same crystal as those in Figure 1c,d but rotated 90° relative to the surface normal (*c* axis). Here the in-plane wavevector of the excitation laser is along the *a* axis ($k_{in} \, // \, a$). Interestingly, we found no interference oscillations in the interior of the crystal as those seen in Figure 1d, indicating that no waveguide EPs are propagating along the *a* axis. To further explore the anisotropic polaritonic responses, we performed energy-dependent s-SNOM imaging. The results are shown in Figure 2, where we plot the s-SNOM imaging data with $k_{in}$ along both the *b* axis

(Figure 2a-e) and $a$ axis (Figure 2f-j) at various excitation energies. Again, we focus on the 1D oscillations at the crystal center (along the direction of the white arrows) that evolve systematically with $E$. For $k_{in}$ // $b$, the oscillations are clearly seen for photon energies from 1.29 to 1.48 eV, and their periods decrease with increasing energy. In the case of $k_{in}$ // $a$, the interference oscillations appear only at energies below 1.32 eV, and there are no clear 1D oscillations from 1.38 to 1.48 eV.

The s-SNOM imaging data shown in Figures 1 and 2 provide direct evidence of in-plane anisotropic EPs of SnS in the near-IR region. To support the experimental data, we performed finite-element simulations of the waveguide EPs using Comsol Multiphysics. In the model, we placed a vertically polarized excitation dipole ($p_z$) right above the sample surface. The optical constants of SnS and $SnS_2$ were obtained from the literature.[22,23,38,39] A detailed description of the Comsol model is given in Supporting Information. The simulation results are shown in Figure 2k-o and Figure S5, where we plot the real-space images of polariton field amplitude ($|E_z|$) and polariton field ($E_z$) of EPs respectively. Here the EPs are launched by a vertically polarized dipole ($p_z$) located at the center of the image. At $E = 1.29$ eV (Figure 2k), the dipole-launched anisotropic EPs propagate at all directions with elliptic wavefronts. As $E$ increases to 1.32 eV (Figure 2l), the EPs show a faster decay along the $a$ axis while keeping a relatively long propagation distance along the $b$ axis. The propagation along the $a$ axis is even shorter at higher energies $E \geq 1.38$ eV (Figure 2m-o). As a result, the EPs appear to be quasi-1D along the $b$ axis. The polaritonic simulations are consistent with s-SNOM imaging data in Figures 1 and 2.

With the s-SNOM imaging data and Comsol simulation results, we were able to perform a quantitative analysis of the dispersion and propagation properties of the anisotropic EPs. In Figure 3a,c, we plot the line profiles extracted across the 1D interference oscillations in the energy-dependent s-SNOM images (Figure 2). We then performed Fourier transforms (FTs) of these profiles to accurately obtain the periods ($\rho$) of the interference oscillations that are linked to the polariton wavelength ($\lambda_p$) in the following relationship:[29,30]

$$\lambda_0/\rho \equiv k_\rho/k_0 \approx \lambda_0/\lambda_p - \cos\alpha. \tag{1}$$

Here, $\lambda_0$ is the excitation photon wavelength, $k_0 = 2\pi/\lambda_0$ is the free-space photon wavevector, $k_\rho = 2\pi/\rho$ is the inverse period of the interference oscillations, and $\alpha \approx 30°$ is the incidence angle of the laser beam relative to the sample plane (see Figure 1a). The FT profiles for $k_{in}$ // $b$ and $k_{in}$ // $a$ are shown respectively in Figures 3b and 3d, where the peaks (marked with blue arrows) correspond to $k_\rho$. We then determined the polariton wavevector ($k_p = 2\pi/\lambda_p$) using Eq. (1) for every given excitation energy, based on which we obtain the energy-momentum dispersion relations of the EPs.

The experimental dispersion data points of EPs obtained through FT analysis (Figure 3b,d) are plotted in Figure 4a,b as black squares, which are sitting on the theoretical dispersion colormaps. In the colormaps, we plot the imaginary part of the reflection coefficients $Im(r_p)$ that represents the photonic density of states (see Supporting Information). Here the TM waveguide modes are visualized as bright curves (marked with blue dashed curves).[29] In addition to the dispersion relations, the $Im(r_p)$ colormaps also reveal the mode broadening ($\Delta k$) that corresponds to the damping (see discussions in the following paragraph). This method of dispersion calculation has been widely used in the studies of polaritons in a variety of materials.[29,30,40,41] In the dispersion diagrams, we also plot the dispersion data points extracted from Comsol simulations (Figure 2k-o and Figure S5). The dispersion relations of the EPs from experimental data, Comsol simulations, and the $Im(r_p)$ colormaps are consistent with each other, which validates our experimental and theoretical approaches. From the dispersion diagrams, we can examine the light-exciton interactions close to the exciton energies $E_a$ and $E_b$ (marked with white dashed lines in Figure 4a,b). The waveguide mode along the $b$ axis exhibits a clear back-bending behavior that is a signature behavior of light-exciton interactions.[28,29] By fitting the dispersion with the coupled oscillator model, we were able to determine the Rabi splitting energy (~160 meV), which is larger than the average polariton linewidth (~105 meV) (see Supporting Information). Therefore, the EPs along the $b$ axis are in the strong coupling regime. The mode coupling is much weaker along the $a$ axis, likely due to the small exciton binding energy. According to literature,[22] excitons along the *b* axis are robust with a binding energy of ~ 50 meV. The binding energy of excitons along the *a* axis, on the other hand, is much smaller and $E_a$ is close to the fundamental bandgap.[22] The light-exciton coupling is much stronger at lower temperatures (e.g., $T$ = 27 K) with more prominent mode bending features close to the exciton energies (Figure S8).

In addition to the polariton dispersion, we also extracted the propagation lengths ($L_{ep}$) of the EPs ($L_{ep} \equiv 1/[2\text{Im}(k_{ep})]$). The extraction was done by fitting the decay trend of the polariton oscillations from both the s-SNOM data (Figure 2a-j) and Comsol simulations (Figure 2k-o and Figure S5). A detailed description of the fitting procedures is given in the Supporting Information. As shown in Figure 4c,d, $L_{ep}$ along both the *a* and *b* axes are over 3 µm at $E$ = 1.29 eV. With increasing $E$, $L_{ep}$ drops systematically along both directions, but the drop along the *a* axis is much faster. As $E$ approaches $E_a \approx 1.39$ eV, $L_{ep}$ along the *a* axis drops below 1 µm and becomes unmeasurable. $L_{ep}$ along the *b* axis, on the other hand, is as high as 2.5 µm at $E$ = 1.38 eV, where quasi-1D EPs were observed (Figure 1d,f). $L_{ep}$ drops to 1 µm or below along the *b* axis when $E$ gets close to $E_b \approx 1.66$ eV. The energy dependence of $L_{ep}$ is fully consistent with the mode broadening behaviors shown in the theoretical dispersion colormaps in Figure 4. The larger the polariton width ($\Delta k$), the smaller the propagation length.

Finally, we explored the dependence of EPs on the thicknesses of SnS crystals. In Figure 5a, we plot the nano-optical images of SnS microcrystals with various thicknesses (*d*) taken at an excitation energy of $E$ = 1.38 eV. Due to the strong damping of EPs along the *a* axis, we only show in Figure 5 the data images with the excitation along the *b* axis ($k_{in}$ // *b*). We focus on the 1D interference oscillations (Figures 1 and 2), which evolve systematically with varying thicknesses. Figure 5b plots the line profiles taken directly across the 1D oscillations in Figure 5a. Using Fourier transform and Eq. (1), we extracted the polariton wavelengths $\lambda_p$ at different sample thicknesses, which are plotted in Figure 5c. Here one can see that $\lambda_p$ decreases systematically with increasing *d*, which is expected since the crystal thickness determines both the out-of-plane ($k_z \sim 1/d$) and in-plane wavevectors of the waveguide mode.[42] For samples with thicknesses over 150 nm, $\lambda_p$ drops below 300 nm that is 3 times smaller than the photon wavelength $\lambda_0$ = 900 nm. The mode confinement is comparable to if not better than waveguide EPs in other materials.[29-32]

In summary, we have performed a comprehensive nano-optical study of SnS microcrystals using s-SNOM. We found through near-field imaging that SnS supports waveguide EPs in near-IR, which are sensitively dependent on both the excitation energy and sample thickness. More interestingly, both the dispersion and transport properties of the EPs are strongly anisotropic in the sample plane. In particular, in the energy range from 1.35 to 1.55 eV, the EPs show quasi-1D propagation along the *b* axis, which has not been reported in other natural polaritonic materials. Future studies with a pump-probe s-SNOM setup[31,32] are expected for the exploration of the ultrafast dynamics of anisotropic EPs in SnS. It is also interesting to study TE waveguide EPs of SnS that could be seen in atomically thin crystals, where active tunability of EPs is possible with electrical gating.

The anisotropic EPs discovered here in SnS are promising for a variety of applications. One potential application is low-pass waveguide filters for planar photonic circuits. The cut-off energies of the filters can be chosen by selecting the direction of signals propagating through the waveguide (i.e., along *a* or *b* axes), Another possible application is selective nanophotonic interconnection. A concept device is sketched in Figure S13, where the signal source is connecting two devices through an SnS interconnector. The two ports for the two devices are along the *a* and *b* axes, respectively. When the incident photonic signal (e.g., on or off signals) is at energies of $E \leq 1.29$ eV, EPs can propagate along all directions in SnS, so both devices can receive the signal. When the incident photonic signal is at energies of 1.38 eV $\leq E \leq$ 1.48 eV, EPs propagate only along the *b* axis, so device 2 will not receive the signal. Therefore, the signal interconnection can be controlled selectively by choosing different energies of the incident signals. Such a directional control of the flow of nanophotonic energy and signals cannot be easily realized in isotropic polaritonic materials without complicated nano-fabrications. With the potential controllability or tunability by chemical doping or electrical gating, SnS-based polaritonic devices could play an important role in future planar optics[43] in the technologically important near-IR region.

**Supporting Information**
Nano-optics setup; Determination of the crystal axes of SnS; Interference mechanisms; COMSOL simulations; Dispersion calculations; Low-temperature dispersion of EPs in SnS; Effect of the SnS2 shell on waveguide EPs; Photonic mode at the air/mica interface; Extraction of the propagation lengths of EPs; Coupling strength of EPs

**Corresponding Authors**
Peter Sutter, Email: psutter@unl.edu,
Zhe Fei, Email: zfei@iastate.edu.

**Notes**
The authors declare no competing interests.

**Acknowledgments**
This work is supported by the National Science Foundation under Grant No. DMR-1945560. The nano-optics setup used in the work is supported in part by Ames Laboratory. Ames Laboratory is operated for the U.S. Department of Energy by Iowa State University under Grant No. DE-AC02-07CH11358. Materials synthesis, electron microscopy, and complementary cathodoluminescence spectroscopy by E.S. and P.S. were supported by the National Science Foundation, Division of Materials Research, Solid State and Materials Chemistry Program under Grant No. DMR-1904843. H.Z. and X.W. are grateful for the support from National Science Foundation under Grant No. CBET-1930866 and CMMI-203246.

**Figure captions**

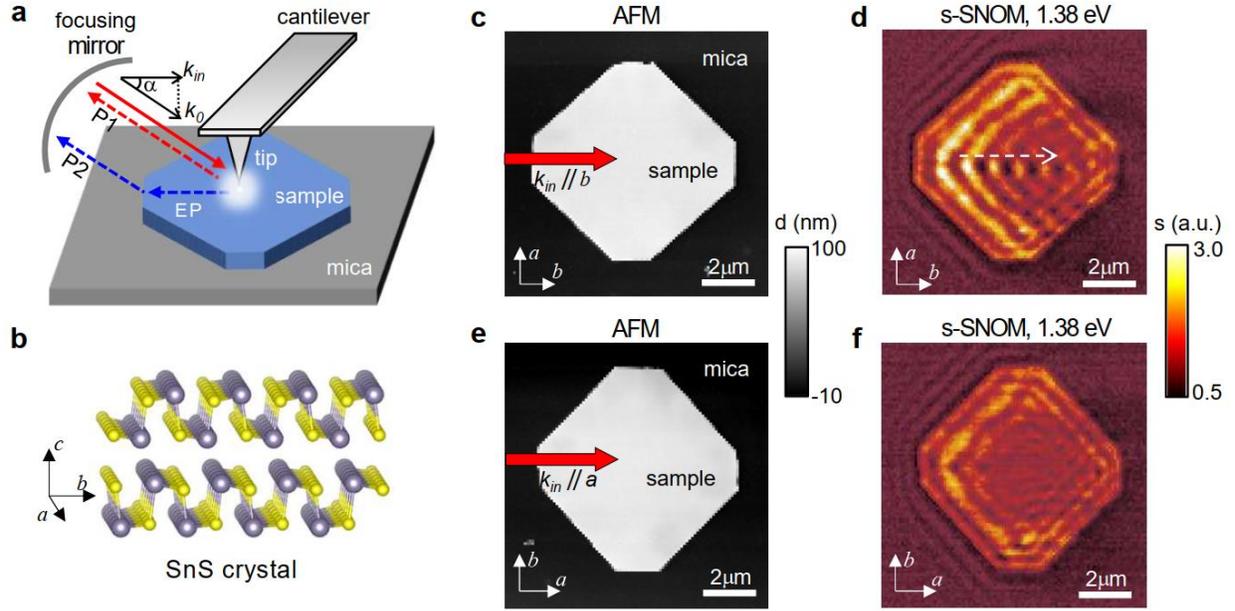

**Figure 1.** (a) Illustration of the experimental setup and the two beam paths (labeled as 'P1' and 'P2') responsible for the formation of the observed interference oscillations. (b) Sketch the crystal structure of SnS. The Sn and S atoms are in silver and yellow, respectively. (c)-(f) The AFM topography and the simultaneously-taken nano-IR images of an SnS microcrystal (thickness = 100 nm) with in-plane wavevector ($k_{//}$) of the excitation laser along the $b$ (c,d) and $a$ (e,f) axes, respectively. The red arrows in (c) and (e) mark the direction of $k_{//}$. The white arrow in (d) marks the 1D oscillations discussed in the main text.

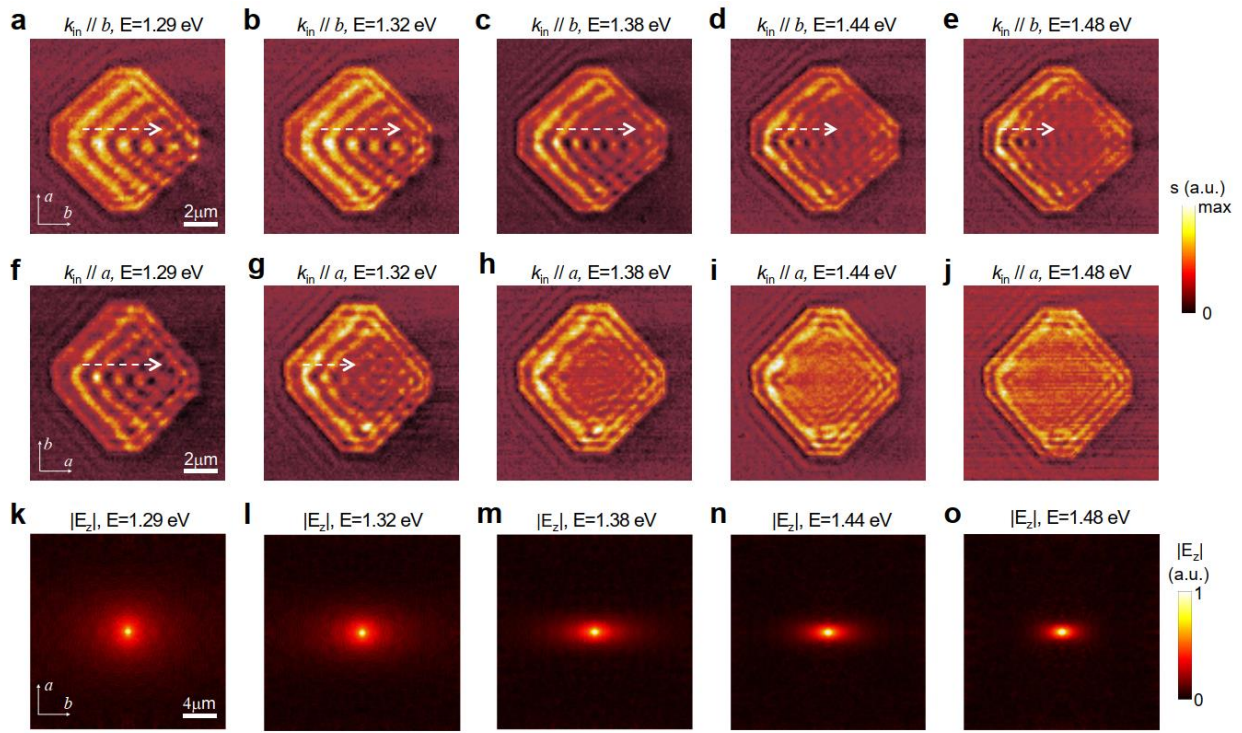

**Figure 2.** (a)-(e) Energy-dependent imaging data of EPs in SnS with the in-plane laser wavevector $k_{in}$ along the $b$ axis. (f)-(j) Energy-dependent imaging data of EPs in SnS with $k_{in}$ along the $a$ axis. Here the sample

is the 100-nm-thick SnS microcrystal shown in Figure 1. (k)-(o) Simulated polariton field amplitude ($|E_z|$) maps of waveguide EPs in 100-nm-thick SnS at various energies. The EPs are excited by a point dipole ($p_z$) located at the center of the image right above the sample surface.

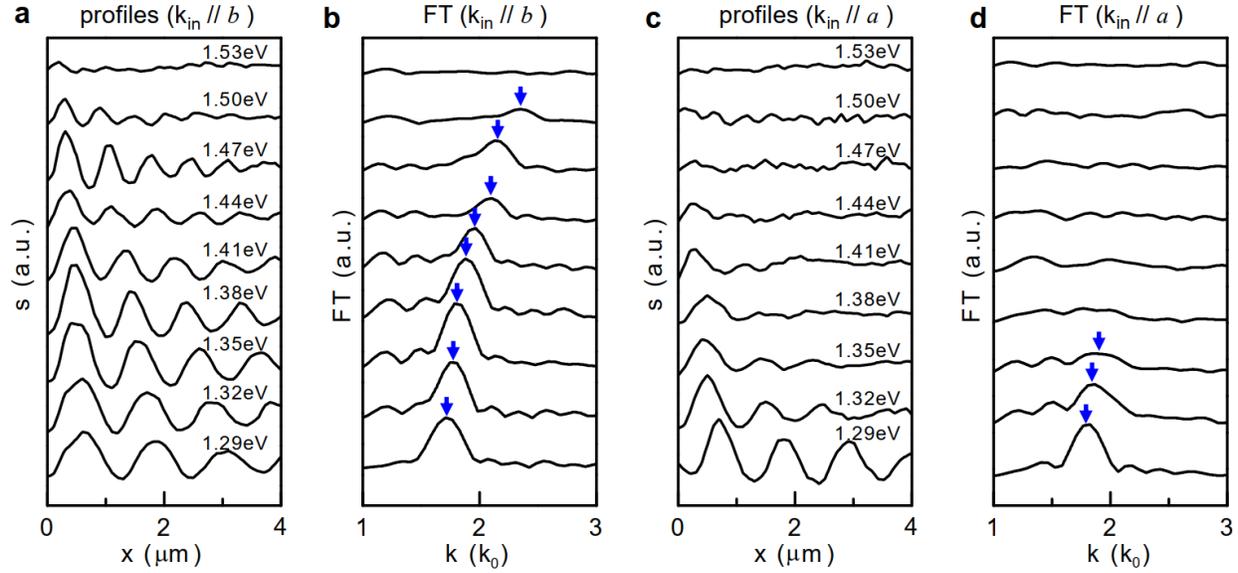

**Figure 3.** Real-space line profiles along the direction of the 1D interference oscillations in Figure 2 and their Fourier-transformed (FT) profiles with the in-plane laser wavevector $k_{in}$ along both the $b$ axis (a,b) and the $a$ axis (c,d), respectively. The unit for the horizontal axes of the FT profiles is $k_0 = 2\pi/\lambda_0$.

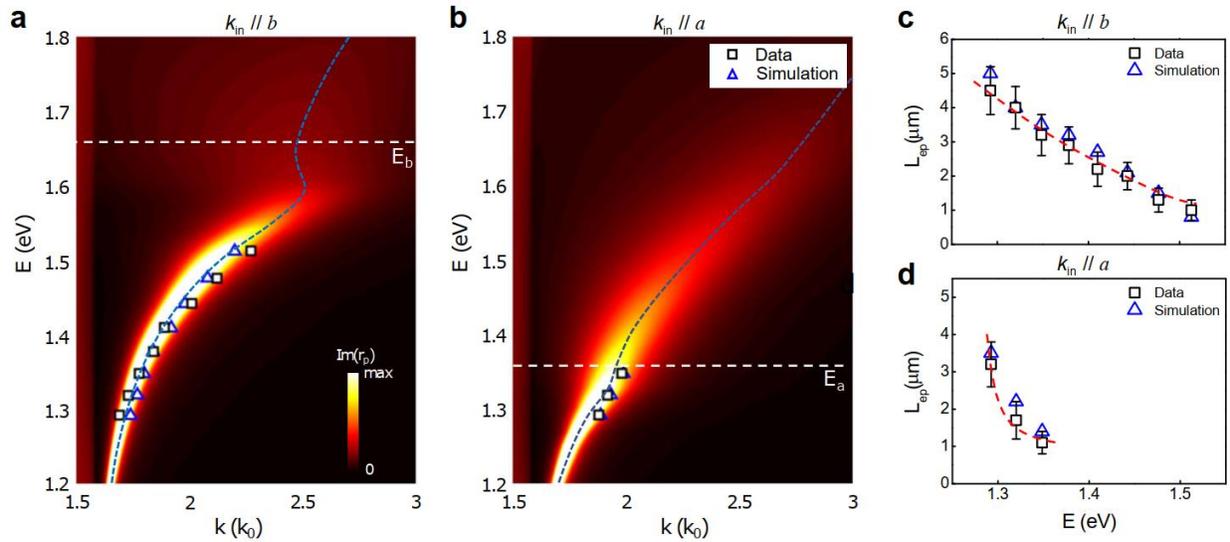

**Figure 4.** (a),(b) Dispersion diagrams of the EPs along the $b$ and $a$ axes, respectively. The colormaps plot the imaginary part of the reflection coefficient $\mathrm{Im}(r_p)$ that represents the photonic density of states. The blue dashed curves mark the dispersion of the waveguide EPs. (c),(d) The propagation lengths of EPs along both $b$ axis and the $a$ axis, respectively. The red curves are drawn to guide the eye. The data points in all panels were obtained from s-SNOM imaging data (squares) and Comsol simulations (triangles).

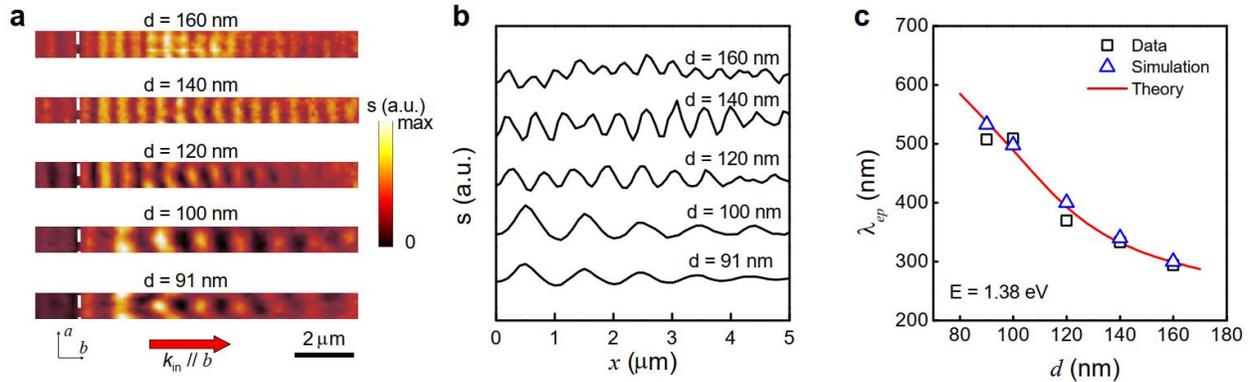

**Figure 5.** (a) Nano-optical images of SnS microcrystals with various thicknesses. Here the excitation energy $E$ = 1.38 eV and the in-plane wavevector is along the $b$ axis. (b) Line profiles taken perpendicular to the interference oscillations in (a). (c) Extracted polariton wavelength $\lambda_p$ versus the thickness of SnS crystals.

## Imaging anisotropic waveguide exciton polaritons in tin sulfide

Yilong Luan[1,2], Hamidreza Zobeiri[3], Xinwei Wang[3], Eli Sutter[4,5], Peter Sutter[6]*, Zhe Fei[1,2]*

[1]Department of Physics and Astronomy, Iowa State University, Ames, Iowa 50011, USA
[2]Ames Laboratory, U. S. Department of Energy, Iowa State University, Ames, Iowa 50011, USA
[3]Department of Mechanical Engineering, Iowa State University, Ames, IA 50011, USA
[4]Department of Mechanical and Materials Engineering, University of Nebraska-Lincoln, Lincoln, NE 68588, USA
[5]Nebraska Center for Materials and Nanoscience, University of Nebraska-Lincoln, Lincoln, NE 68588, USA.
[6]Department of Electrical and Computer Engineering, University of Nebraska-Lincoln, Lincoln, NE 68588, USA

*Corresponding to: (P.S.) psutter@unl.edu, (Z.F.) zfei@iastate.edu.

**List of contents:**

## 1. Nano-optics setup

To image the propagative waveguide EPs in SnS, we applied the s-SNOM from Neaspec GmbH The s-SNOM was built based on a tapping-mode Atomic Force Microscope (AFM). The AFM tips used in the study were Pt/Ir-coated silicon tips (Arrow NCPT from Nanoandmore GmbH) with a tapping frequency of ~270 kHz. The tapping amplitude of the tip was set to be about 50 nm. For optical excitations, we used a Ti:sapphire laser (Spectra-Physics, Tsunami) operating at the continuous-wave mode with a photon energy tunable from 1.3 to 1.8 eV that covers the exciton and bandgap energies of SnS. The main observable of the s-SNOM is the complex near-field scattering signal that is modulated due to the tip tapping. Demodulating the signal at the $n^{th}$ harmonics ($n \geq 2$) of the tapping frequency can effectively suppress the background signal ($n = 2$ in the current work). In addition, the pseudo-heterodyne interferometric detection method is used to extract both the amplitude ($s$) and phase ($\psi$) of the near-field scattering signal. In the current work, we mainly discuss the amplitude part of the signal that is sufficient for describing propagative EPs. All s-SNOM measurements were performed at ambient conditions.

## 2. Determination of the crystal axes of SnS

As introduced in the main text, our samples are SnS microcrystals coated with a thin shell of $SnS_2$. In Figure S1a, we show an optical photo of the sample, where tens of microcrystals are sitting on the mica substrate. Prior to the s-SNOM imaging measurements, we first determined the in-plane crystal axes (i.e., *a* and *b* axes) of SnS. The most convenient way to determine the crystal orientation is by inspecting the crystal shape.[1,2] As shown in Figure S1b, the crystal is slightly elongated along the *b* axis. As a result, the crystal corner angles are 85° and 95°, respectively. Therefore, by measuring the corner angle, we can determine the crystal axes. Alternatively, we also applied polarization-dependent Raman spectroscopy to confirm the crystal axes. Here the incident laser beam is polarization-controlled, and the detector collects photons from all polarizations. Due to the strong anisotropy of SnS, Raman spectra show sensitive dependence on the polarization direction of the incident laser.[1] In Figure S1c, we plot the Raman spectra of a SnS microcrystal with laser polarization along the *a* and *b* axes. When polarization is parallel with the *a* axis (the black curve in Figure S1c), there are two outstanding peaks at 159 cm$^{-1}$ ($B_{3g}$) and 188 cm$^{-1}$ ($A_g$), respectively. When polarization parallel with *b* axis (the red curve in Figure S1c), one more prominent peak at around 93 cm$^{-1}$ ($A_g$) emerges in addition to the two peaks at 159 cm$^{-1}$ and 188 cm$^{-1}$. The Raman peak at 93 cm$^{-1}$ was used to distinguish the *a* and *b* axes of SnS.[1]

## 3. Interference mechanisms

As discussed in the main text, the real-space fringes or oscillations of EPs observed in our s-SNOM imaging data were formed due to the interference between tip-back-scattered photons and edge-scattered

EPs (termed as "Mechanism I"). In the main text, we focus on the discussions of the string of 1D oscillations at the center of the SnS microcrystal. The optical paths responsible for the formation of these interference oscillations are sketched in Figure 1a of the main text and Figure S2a. In this case, the incident laser beam is perpendicular to the short edge (labeled as edge I in Figure S2a) and has an incident angle of $\alpha \approx 30°$ relative to the sample plane. Upon tip illumination, part of the laser beam is scattered directly by the tip to the detector (labeled as path P1). The tip also excites in-plane EPs and propagate perpendicular to the short edge along the *a* axis (e.g., *edge I* in Figure S2a). When reaching *edge I*, the EPs are scattered to be photons (labeled as path P2) that are collected by the detector. The photons collected from the two paths interfere with each other and thus generating the 1D interference oscillations at the crystal center. In addition to these 1D oscillations, there are also interference fringes parallel to the relatively long edges (e.g., *edge II* in Figure S2b) that are about 43° relative to the *b* axis. The general mechanism for the fringe formation is similar to those of the 1D oscillations. The main difference is that the EPs responsible for the fringe formation parallel to *edge II* are propagating along a direction off the two principal crystal axes (i.e., *a* and *b* axes). With careful boundary-condition analysis, we found that the propagation direction of the EPs has an angle of $\phi \approx 20°$ relative to the *b* axis to form interference fringes parallel to *edge II* (see Figure S2b).

In addition to "mechanism I" discussed above and in the main text, there are also other possible interference mechanisms. One possible mechanism (termed as "mechanism II") is related to edge excitation of EPs followed by tip scattering, and the other involves tip excitation of EPs followed by edge reflection and tip scattering (termed as "mechanism III"). As discussed below, both mechanisms II and III are not responsible for the interference oscillations/fringes in the current work.

Mechanism II (edge excitation → tip scattering) requires that the focused laser spot (radius ~ 1 μm) is at the sample edge, which is only possible when the tip is very close to the sample edge because the laser is always focused on the tip apex. Therefore, for interference fringes or oscillations 1 μm away from the sample edge, edge excitation has little contributions. In addition, edge excitation followed by tip scattering is exactly the reverse process of tip excitation followed by edge scattering, so the distance of the optical paths and hence the interference fringes are expected to be the same in the two cases. Finally, the s-SNOM tip is in principle more efficient in polariton excitation than the sample edge due to its metallicity. Considering the above three factors, we believe Mechanism II is not responsible for the interference fringes or oscillations observed in our work.

Mechanism III (tip excitation → edge reflection → tip scattering) plays important role when polaritons or plasmons are highly confined (confinement factor $k_p/k_0 \gg 1$), which is typical for plasmons and polaritons in the mid-infrared region (e.g., graphene plasmons).[3,4] Here, highly confined polaritons or plasmons are efficiently reflected at the sample edge due to the large impedance mismatch. In the case of polaritons or plasmons in the near infrared or visible range (e.g., metal plasmons or exciton polaritons), the confinement is weaker. As a result, only a small portion of polaritons or plasmons can be reflected. Take EPs of SnS for example, the confinement factor $k_p/k_0$ is in the range of 1.6-2.5 (see Figure 4a in the main text), therefore the polariton reflectance at the sample edge $R \approx |(k_p - k_0)/(k_p + k_0)|^2$ is in the range of 5% to 18%. Moreover, additional geometric and intrinsic damping during the round-trip propagation (from the tip to the edge and back to the tip) further weakens the reflected EPs. Therefore, Mechanism III also does not play an important role in the fringes/oscillations observed in SnS.

## 4. COMSOL simulations

To support the experimental study, we performed finite-element simulations of waveguide EPs in SnS with COMSOL Multiphysics. To excite EPs, we placed a *z*-polarized point dipole ($p_z$) right above the sample surface. We used two types of models to simulate the SnS microcrystal sample. The first one is a realistic model, where the sample was set to be a four-layer heterostructure ($SnS_2$/SnS/$SnS_2$/mica). Due to the ultra-thin $SnS_2$ shell layer (thickness ~ 3 nm), the realistic model requires ultra-fine messing, so it is time-consuming and not suitable for the simulations of large samples. The second one is an effective model, where the SnS layer is set to be in a homogeneous dielectric environment. The effective permittivity ($\varepsilon_{eff}$) of the homogeneous dielectric environment can be considered as an average value of air, $SnS_2$, and mica.

We treated $\varepsilon_{eff}$ as a fitting parameter that was determined by comparing the two models. As shown in Figure S3, the simulated polariton field (out-of-plane $E_z$ field) maps of the two models are consistent with each other when using $\varepsilon_{eff} = 2.2$ for the 100-nm-thick microcrystal sample. The consistency is also checked at all other excitation energies. Due to the simplicity of the effective model, the simulations are much more efficient. Moreover, we can simulate very large samples with a size of tens of microns, which is necessary for the extraction of the propagation lengths of EPs. The main simulation results in this work were produced with the effective model. The permittivity of SnS along the *a* and *b* axes (plotted in Figure S4) used in the Comsol simulations and dispersion calculations is from previous literature.[5,6] The *c*-axis permittivity of SnS is adopted from Ref. 7. In Figure S4, we also mark the optical bandgap or exciton energies (blue arrows), which are 1.66 eV and 1.39 eV along the *b* and *a* axes, respectively. The permittivity of $SnS_2$ is set to be about 10 in the *ab* plane and 6 along the *c* axis in our spectral range.[8,9] The permittivity of mica is set to be 2.5 according to Ref. 10.

In Figure 2k-o and Figure S5a-e, we plot respectively the polariton field amplitude ($|E_z|$) and polariton field ($E_z$) maps at various excitation energies. Based on these field maps, we were able to determine the polariton wavelength and propagation length, which match well the experimental results (see Figure 4, Figure S6 and Figure S7). To better visualize the anisotropy of the EP modes, we performed 2D Fourier transform of the $E_z$ field maps in Figure S5a-e to generate the isofrequency contours. The results are shown in Figure S5f-j, where one can see that the EP mode has a clear elliptic shape at $E = 1.29$ eV and $E = 1.32$ eV. As $E$ increases to 1.38 eV and above, the top part of the ellipses (i.e., corresponding to the mode along the *a* axis) becomes strongly weakened due to the high damping, so EPs prefer propagating along the *b* axis.

## 5. Dispersion calculations

In Figure 4 of the main text, we plot the dispersion diagrams of SnS along both the *a* and *b* axes, where the data points obtained from s-SNOM experiments and Comsol simulations are overlaid on the theoretical dispersion colormaps. In the energy-momentum dispersion colormaps, we plot the imaginary part of the *p*-polarization reflection coefficient Im($r_p$), which represents the photonic density of states. The bright curves shown in the dispersion colormaps correspond to transverse magnetic (TM) waveguide modes.[11] The transverse-electric waveguide modes, on the other hand, can be revealed when plotting the imaginary part of the *s*-polarization reflection coefficient Im($r_s$).[12] In the transfer-matrix calculations, we considered the entire $SnS_2$/SnS/$SnS_2$/Mica heterostructure. Take the microcrystal sample in Figs. 1 and 2 in the main text for example, the crystal thickness is ~100 nm. Considering a 3-nm-thick $SnS_2$ shell at the top and bottom,[1] the SnS core has a thickness of ~94 nm.

## 6. Low-temperature dispersion of EPs in SnS

To explore theoretically the low-temperature behavior of waveguide EPs in SnS, we plot in Figure S8a,b the calculated polariton dispersion of the 100-nm-thick SnS microcrystal at $T = 27$ K. The low-temperature permittivity of SnS is from Refs. 5 and 6. For comparison, we plot in Figure S8c,d the dispersion diagrams of EPs at $T = 300$ K (replotted from Figure 4 in the main text). Compared to room-temperature dispersion color plots, the waveguide polariton mode at $T = 27$ K (Figure S8) is sharper due to the smaller damping at the lower temperature. Besides, the exciton energies slightly increase at the lower temperature. Similar temperature dependence of exciton energies has also been seen in other van der Waals semiconductors.[13,14] Furthermore, the light-exciton coupling is much stronger at $T = 27$ K with more prominent mode bending features close to the exciton energies.

## 7. Effect of the $SnS_2$ shell on waveguide EPs

As discussed in the main text, the SnS microcrystals studied in this work are coated with a thin $SnS_2$ shell that has a thickness of ~ 3 nm at the top and bottom surfaces.[1] Here we wish to evaluate the effect of the thin $SnS_2$ shell on the waveguide EPs. In Figure S9, we plot the calculated dispersion diagrams of waveguide EPs along the two principal axes of SnS with (Figure S9a,c) and without (Figure S9b,d) consideration of the $SnS_2$ shell. From Figure S9, one can see that the $SnS_2$ shell has a very limited effect on the EPs. Close examination indicates that the SnS$_2$ shell only induces a slight (~3-4%) decrease of polariton wavelengths of EPs propagating along both the *a* and *b* axes. The polaritonic anisotropy along the *a* and *b* axes, on the other hand, is solely due to the SnS core.

## 8. Photonic mode at the air/mica interface

The fringes are also seen on the mica substrate (Figures 1,2 and Figure S10a), which are generated due to the interference of photonic modes propagating at or close to the surface of the mica substrate. The interference mechanism for the substrate fringes is sketched in Figure S10b. To verify that, we performed a dispersion analysis of the substrate mode. In Figure S10c, we show the excitation-energy-dependent fringe profiles extracted along the blue dashed line in Figure S10a. With Fourier transform (Figure S10d), we were able to determine the fringe period, which can be converted directly into the mode wavevector of the substrate $k_s$ using the following equation:

$$\lambda_0/\rho \equiv k_s/k_0 \approx \lambda_0/\lambda_s + \cos\alpha \equiv k_s/k_0 + \cos\alpha. \tag{S1}$$

Note the difference between Eq. S1 with Eq. 1 in the manuscript ('+' sign instead of '-' sign). Following the Fourier transform, we obtained the dispersion data points, which match well the theoretical dispersion colormap (Figure S10e). From the dispersion diagram, one can see that the wavevector of the substrate mode is roughly proportional to the free-space photon wavevector $k_0$ indicating their photonic nature (note that the *k* axis in the dispersion diagram is normalized to $k_0$). The mode wavevector is between $k_0$ to $nk_0$, where $n \approx 1.6$ is the refractive index of mica. Therefore, we believe the substrate mode measured here corresponds to in-plane photons at the air/mica interface.

## 9. Extraction of the propagation lengths of EPs

In this section, we describe the extraction processes of the propagation lengths $L_{ep} \equiv 1/(2q_2)$, where $q_2$ is the imaginary component of the polariton wavevector $q = q_1 + iq_2$. We extracted $L_{ep}$ from both the experimental data and COMSOL simulations. The experimental $L_{ep}$ was extracted from the polariton fringe profiles shown in Figure 3a,c in the main text. We first subtracted the baseline signal of the sample to obtain the pure EP fringe oscillations. The baseline signal comes mainly from the background signal of the sample without the generation of the propagative EPs. Detailed introductions about baseline subtraction can be found in Ref. 12. The baseline-corrected fringe profiles are plotted as black curves in Figs. S6a and S7a, which show clear decay with distance. We then performed envelop fitting of the profiles with a radial exponential decay function $x^{-1/2}\exp(-x/2L_{ep})$, which is expected for radially propagating 2D waves. Note that not all experimental profiles can be fitted due to the high damping. As shown in Figs. S6a, the profile at $E = 1.54$ eV cannot be fitted for $k_{in} // b$. In the case of $k_{in} // a$ (Figure S7a), the profiles at $E \geq 1.38$ eV cannot be fitted. Similar fitting procedures were also applied to extract $L_{ep}$ from the simulated EP oscillations (see Figs. S6b and S7b).

## 10. Coupling strength of EPs

In this section, we estimate the coupling strength of EPs propagating along the *b* axis of SnS. The criterion for strong coupling is that the Rabi splitting energy $\Omega_R \approx 2hg$ is larger than the average EP linewidth $(\Gamma_{ex} + \Gamma_{ph})/2$, where $hg$ is the coupling energy, $\Gamma_{ex}$ and $\Gamma_{ph}$ are the linewidths (full width at half maximum) of exciton and waveguide photon mode.[15] To determine the Rabi splitting energy, we fit the dispersion relationship of the EP mode along the *b* axis of SnS using the equation below:[15,16]

$$E_\pm \approx \frac{E_{ph} + E_{ex}}{2} \pm \frac{1}{2}\sqrt{(E_{ph} - E_{ex})^2 + 4(hg)^2}. \tag{S2}$$

Note that the fitting is mainly based on the bottom-branch of the EP mode that was verified experimentally. Similar approach has been adopted in Ref. 16. The fitting result is shown in Figure S11, where the fitting curves with Eq. S2 match well the dispersion relation of EPs revealed by the colormap. Note that the dispersion colormap is a replot of Figure 4a without normalization of the *k* axis to $k_0$. Through the fit, we obtain the Rabi splitting energy to be about 160 meV, which is comparable to or even bigger than those reported in other materials. The exciton linewidth of SnS along the *b* axis is about 140 meV at room temperature by fitting the dielectric function from previous literature (see Figure S12a,b).[5] The linewidth of the bare waveguide photon mode is estimated to be 70 meV at the exciton energy (see Figure S12c), so the average polariton linewidth is about 105 meV, which is smaller than the Rabi splitting energy. Therefore, we conclude that the EPs of SnS along the *b* axis is in the strong coupling regime.

**References for the Supporting Information**

**Supporting Figures**

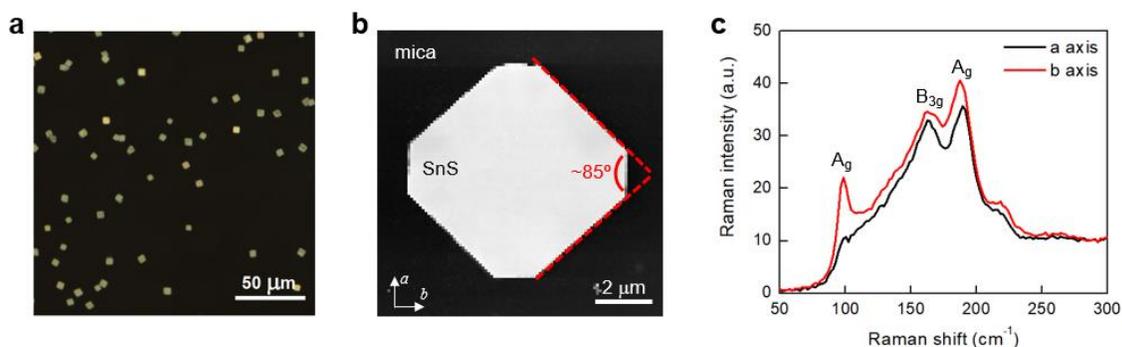

**Figure S1.** (a) A large-area optical photo of SnS microcrystals on a mica substrate. (b) The AFM image of an SnS microcrystal. The marked angle of the crystal corner is 85º. (c) Polarization-dependent Raman spectra of SnS along both the *a* and *b* axes. Detailed discussions are given in Section 2 of the Supporting Information.

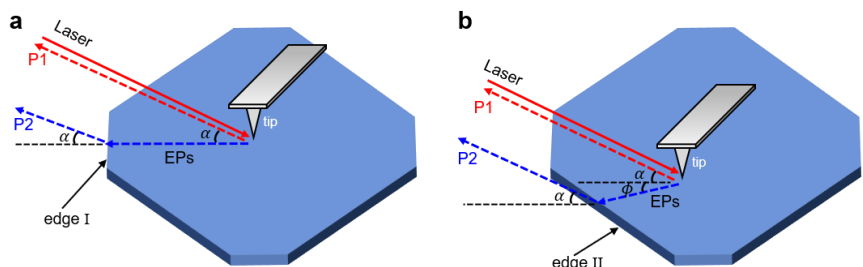

**Figure S2.** Illustrations of the formation mechanism of 1D interference oscillations at the crystal center (a) and interference fringes parallel to the long edges (b). Detailed discussions are given in Section 3 of the Supporting Information.

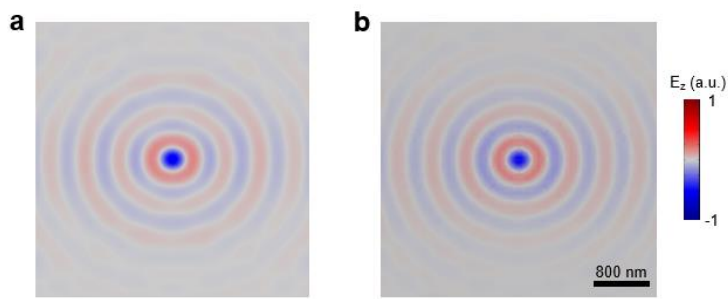

**Figure S3.** The simulated polariton field ($E_z$ field) maps with the realistic model (a) and the effective model (b). Detailed discussions are given in Section 4 of the Supporting Information.

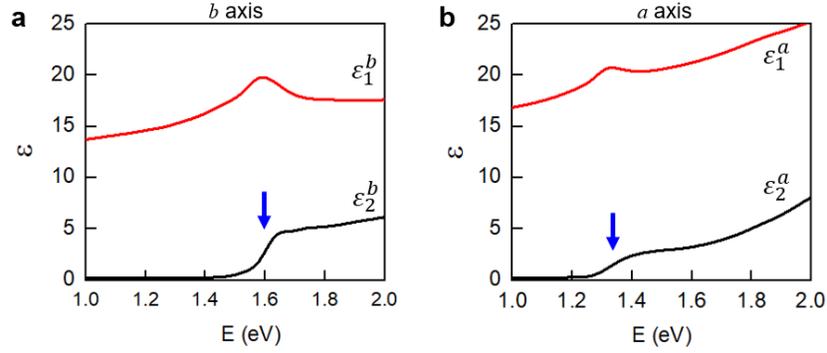

**Figure S4.** The real part (red) and the imaginary part (black) of the permittivity along *b* axis (a) and *a* axis (b). The blue arrow marks the optical bandgap or exciton energy.

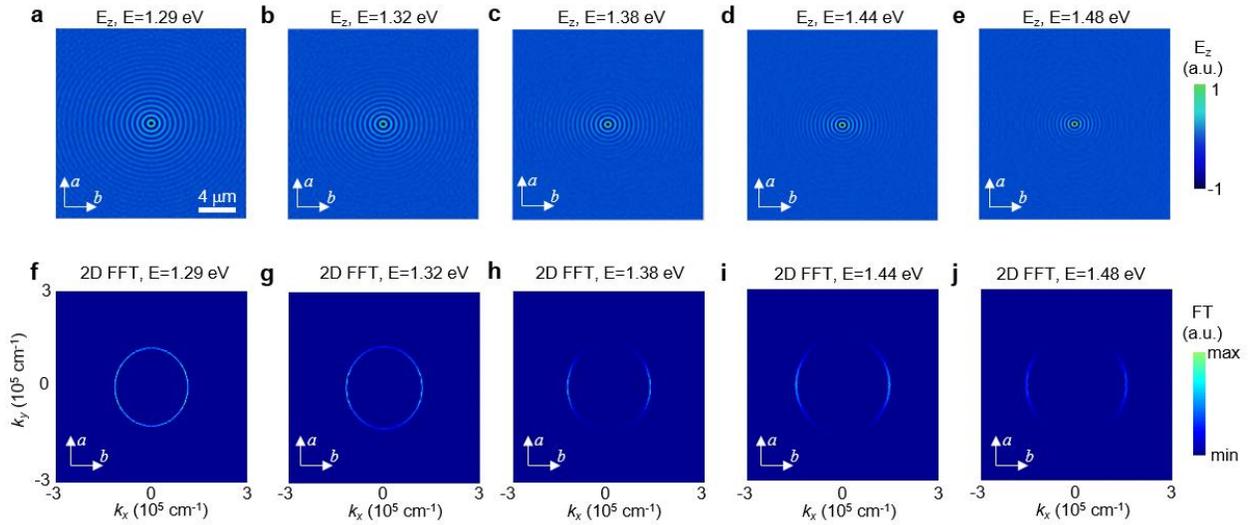

**Figure S5.** (a-e) Simulated out-of-plane field ($E_z$) maps of waveguide EPs in 100-nm-thick SnS at various energies. The EPs are excited by a point dipole ($p_z$) located at the center of the image right above the sample surface. (f-j) Energy-dependent isofrequency contour maps of waveguide EPs generated by the Fourier transform of the $E_z$ maps in (a-e). Detailed discussions are given in Section 4 of the Supporting Information.

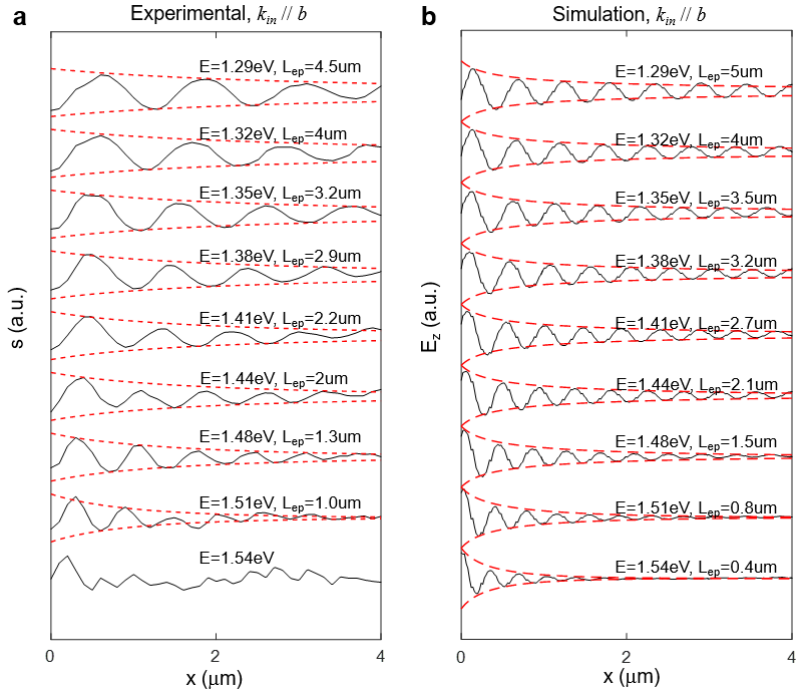

**Figure S6.** Fitting of the propagation length ($L_{ep}$) of EPs along the *b*-axis based on the experimental data (a) and Comsol simulations (b).

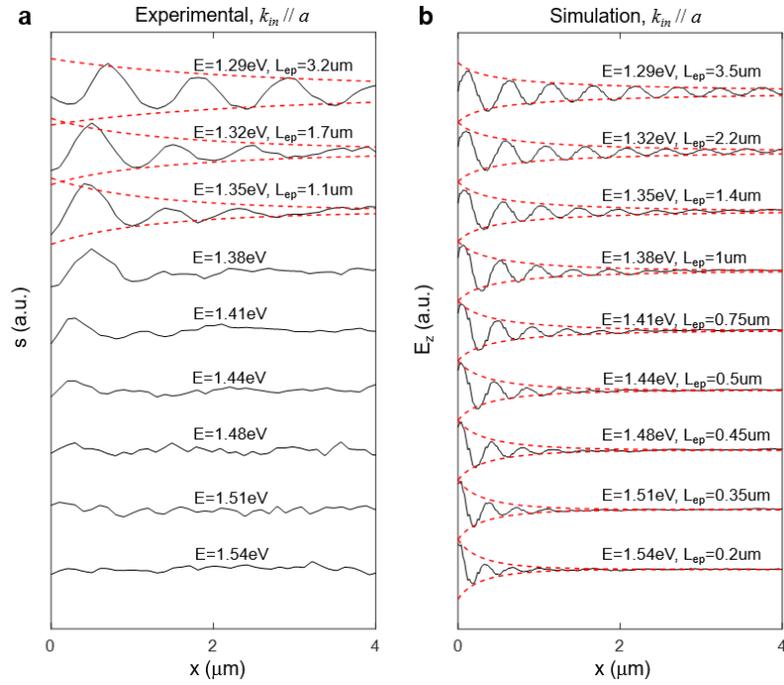

**Figure S7.** Fitting of the propagation length ($L_{ep}$) of EPs along the *a*-axis based on the experimental data (a) and Comsol simulations (b).

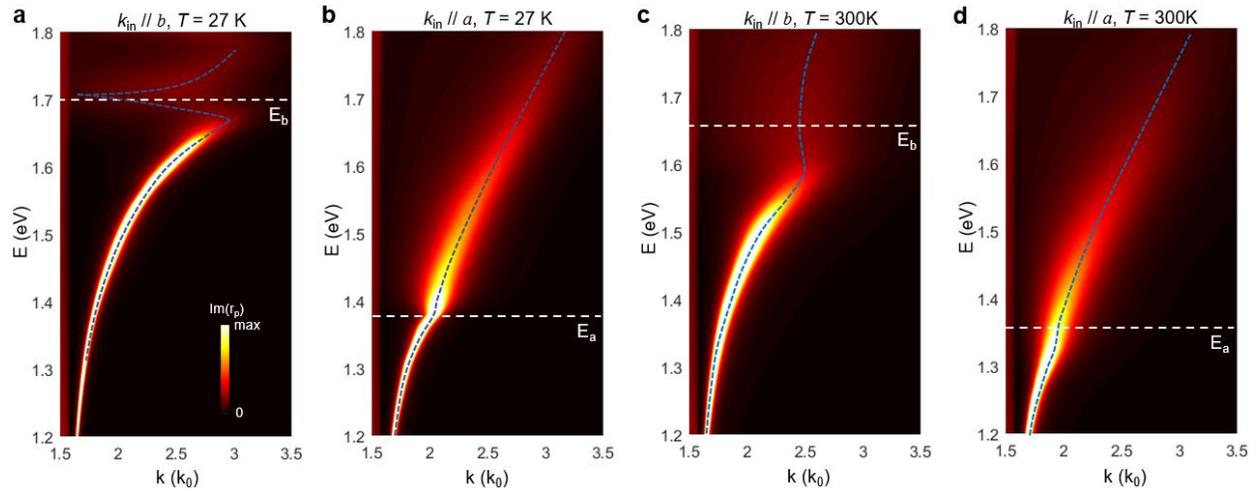

**Figure S8.** The dispersion colormaps of SnS along the two principal axes at $T = 27$ K (a,b) and $T = 300$ K (c,d). Detailed discussions are given in Section 6 of the Supporting Information.

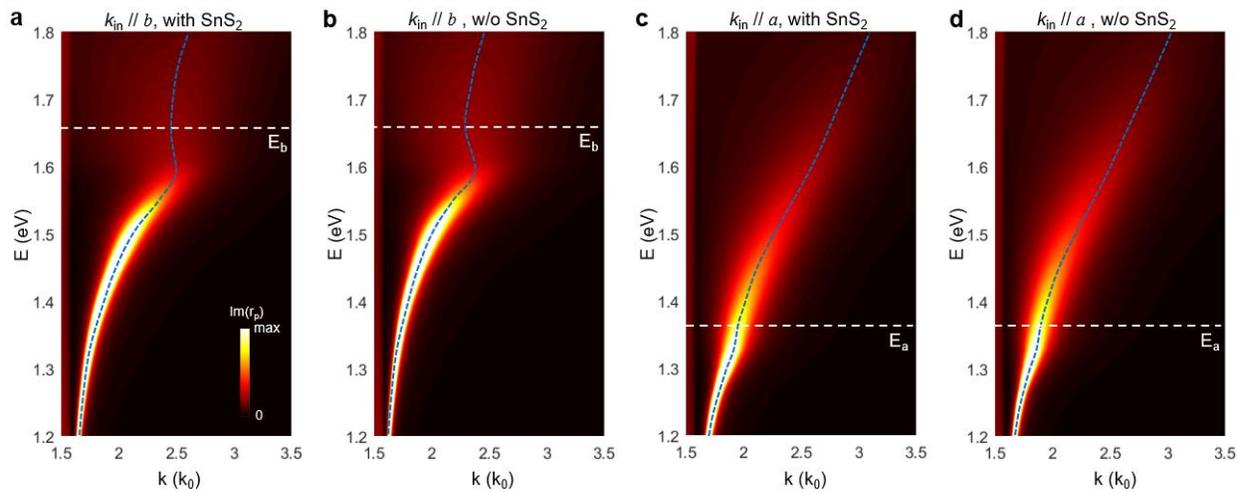

**Figure S9.** The calculated dispersion colormaps of waveguide EPs along the two principal axes of SnS with (a,c) and without (b,d) consideration of the thin $SnS_2$ shell.

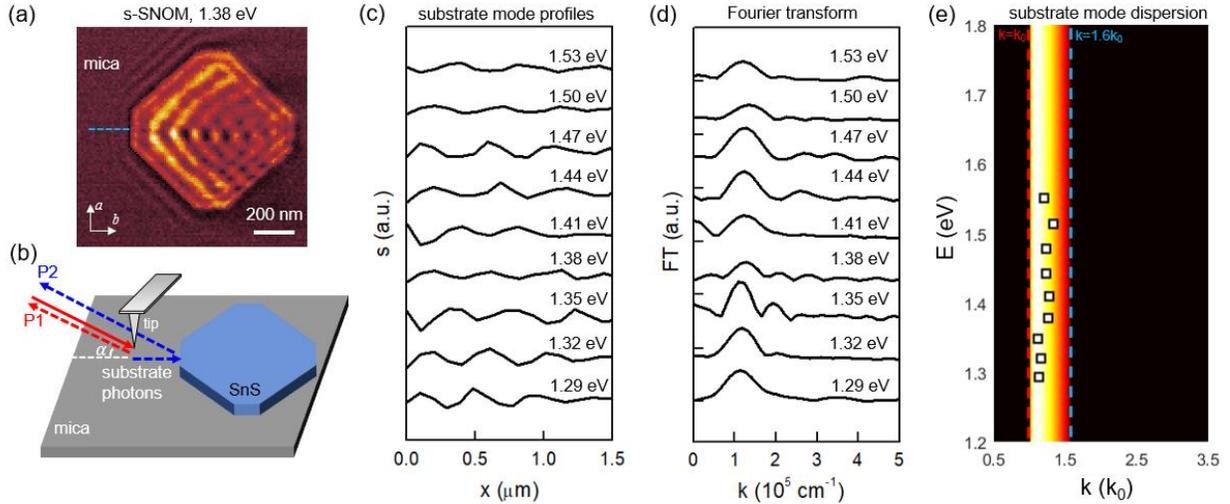

**Figure S10.** (a) Nano-optical imaging data of the SnS crystal on mica substrate at $E = 1.38$ eV. (b) Illustration of the interference mechanism for the formation of fringes on the substrate. (c) Fringe profiles at various excitation energies taken along the blue dashed line shown in (a). (d) Fourier transform of the energy-dependent fringe profiles shown in (c). (e) Experimental and theoretical dispersion diagram of the substrate photon mode of mica. The two vertical dashed lines mark the photon dispersion in air ($k = k_0$) and in bulk mica ($k=1.6k_0$). Detailed discussions are given in Section 8 of the Supporting Information.

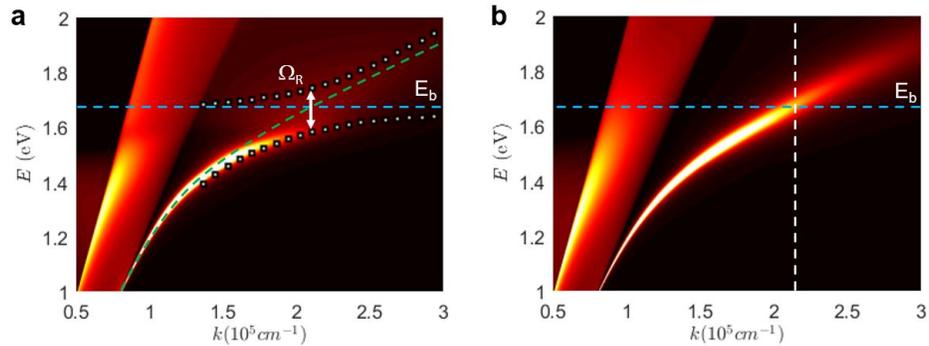

**Figure S11.** (a) Fitting the dispersion relation of the EP mode along the $b$ axis of SnS. The white dotted curves were produced with Eq. S2. (b) Calculated dispersion relation of the bare waveguide photon mode without coupling with excitons along the $b$ axis of SnS. Detailed discussions are given in Section 10 of the Supporting Information.

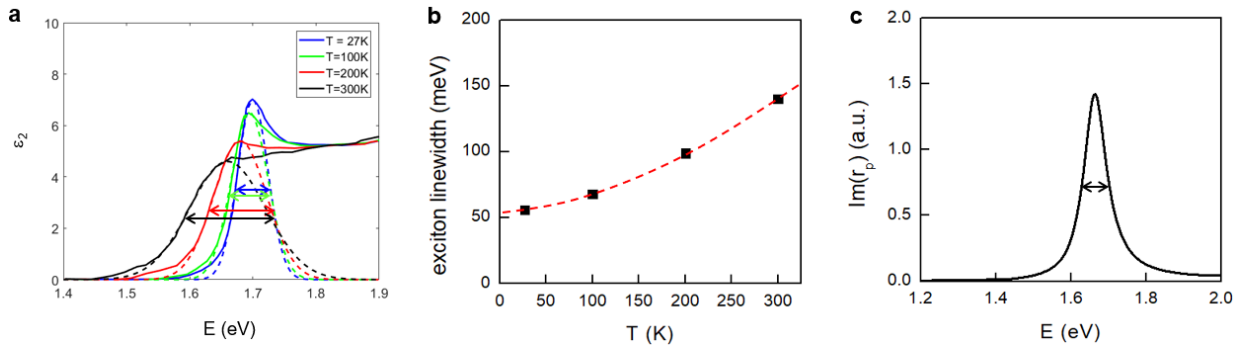

**Figure S12.** (a) Fitting the exciton linewidth from the *b*-axis permittivity of SnS from Ref. 5. (b) Temperature-dependent linewidth of excitons along the *b* axis of SnS based on the fitting in (a). (c) Line profile of the photonic mode taken along the vertical dashed line in Figure S11b. Detailed discussions are given in Section 10 of the Supporting Information.

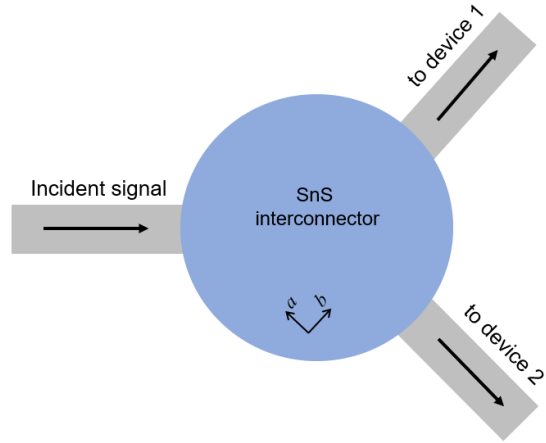

**Figure S13.** Illustration of a selective nanophotonic interconnector based on anisotropic EPs in SnS. When the incident photonic signal is at energies of $E \leq 1.29$ eV, both devices 1 and 2 can receive the signal. When the incident photonic signal is at energies of $E \geq 1.38$ eV, only device 1 can receive the signal.